\documentclass[%
 reprint,
 superscriptaddress,
showpacs, preprintnumbers,
 amsmath,amssymb,
 aps,
]{revtex4-1}

\usepackage{amsmath}
\usepackage{mathtools}
\usepackage{bbold}
\usepackage{braket}
\usepackage{graphicx}
\usepackage{dcolumn}
\usepackage{bm}
\begin{document}

\preprint{APS/123-QED}

\title{Superconductivity-induced features in electronic Raman spectrum of monolayer graphene} 

\author{A. Garc\'{i}a-Ruiz}
\email{agr31@bath.ac.uk}
\affiliation{Department of Physics, University of Bath, Claverton Down, Bath BA2 7AY, United Kingdom}
\author{M.~Mucha-Kruczy\'{n}ski}
\affiliation{Department of Physics, University of Bath, Claverton Down, Bath BA2 7AY, United Kingdom}
\affiliation{Centre for Nanoscience and Nanotechnology, University of Bath, Claverton Down, Bath BA2 7AY, United Kingdom}
\author{V. Fal'ko}
\affiliation{National Graphene Institute, University of Manchester, Booth St E, Manchester, M13 9PL, United Kingdom}
\affiliation{School of Physics and Astronomy, University of Manchester, Oxford Road, Manchester, M13 9PL, United Kingdom}

\date{\today}

\begin{abstract}
Using the continuum model, we investigate theoretically contribution of the low-energy electronic excitations to the Raman spectrum of superconducting monolayer graphene. We consider superconducting phases characterised by an isotropic order parameter in a single valley and find a Raman peak at a shift set by the size of the superconducting gap. The height of this peak is proportional to the square root of the gap and the third power of the Fermi level, and we estimate its quantum efficiency as $I\sim10^{-14}$.

\begin{description}
\item[Usage]
\end{description}
\end{abstract}

\pacs{Valid PACS appear here}
\maketitle

\section{\label{sec:level1}Introduction}
Graphene, a single-layer of carbon atoms arranged in regular hexagons, has been the topic of intense studies since its mechanical isolation in 2004 \cite{graphene_2004}. Its unique properties include a linear electronic dispersion relation, which in turn gives rise to exotic effects like Klein tunnelling \cite{Klein_Paradox}, and a strong electron-photon coupling, which makes observation of free-standing graphene with the naked eye possible \cite{naked_eye}. Recently, a significant effort has been directed towards adding one more entry to the list of graphene properties, namely, superconductivity \cite{Tuning, Heersche1, Heersche2, Natterer, Li-Decorated, Ca-doped, TunnellingSpectroscopy}. This aim, motivated by the idea of combining the physics of Dirac fermions and Cooper pairs, was achieved by the proximity effect  \cite{Tuning, Heersche1, Heersche2, Natterer,TunnellingSpectroscopy} or doping with metallic adatoms \cite{Li-Decorated, Ca-doped}. However, due to the smallness of the gap in the electronic dispersion and the difficulty in preparing good quality samples, direct and unambiguous experimental observation of superconductivity has been a real challenge. Tunnelling spectroscopy measurements \citep{TunnellingSpectroscopy}, angle-resolved photoemission spectroscopy \cite{Li-Decorated} and SQUID magnetometry  \citep{Ca-doped} have been used to confirm the presence of superconductivity in doped-graphene and related systems. Unfortunately, all these methods require sophisticated fabrication processes or equipment, and are usually quite expensive.

On the other hand, Raman scattering is known to be a powerful technique which allows to extract a wealth of information about graphene and other graphene-related materials, such as, number of layers, defect density, doping level or presence of strain in the sample \cite{Versatil_Tool}. In all those cases, the Raman shift equal to the difference between the energies of the incident and detected photons arises because part of the energy is spent on exciting the crystal lattice of graphene. However, purely electronic processes can also give rise to Raman features \cite{Light_Scattering}. In particular, it has been predicted \cite{Theory_ERS_MLG, Kashuba_Falko1, Role_of} and later confirmed experimentally \cite{Exp_ERS_graphene_1, Exp_ERS_graphene_2, ExpConf, ExpConf2, ExpConf3} that in monolayer graphene a Raman process resulting in creation of an electron-hole pair leads to a characteristic linear feature in the Raman spectrum, a consequence of the linear electronic density of states. 

Here, we discuss theoretically the feasibility of using electronic Raman scattering (ERS) to detect superconductivity in graphene. We start by presenting in Section \ref{sec:level2} the theory of the electronic contribution to the Raman scattering for monolayer graphene \cite{Role_of}. Then, in Section \ref{sec:level3}, we introduce a pairing interaction of electrons in graphene, and focus on phases characterised by an isotropic order parameter in a single valley. For these phases, we predict a peak in the Raman spectrum at the Raman shift corresponding to the size of the superconducting gap, and with quantum efficiency proportional to the size of the gap and to the third power of the Fermi level.

\section{\label{sec:level2}ERS in graphene}

In this section, we discuss processes in which inelastic scattering of a photon off monolayer graphene is accompanied by formation of an electron-hole pair within the same electronic band. Following previous work on electronic Raman scattering in graphene materials \cite{Theory_ERS_MLG, Theory_ERS_BLG, Kashuba_Falko1, Role_of}, we evaluate the scattering amplitude corresponding to such processes as well as the resulting angle-resolved probability, spectral density and quantum efficiency. These results form the basis for the analysis of the electronic Raman spectroscopy features arising due to the presence of Cooper pairs.

The electronic properties of graphene are well described by a tight-binding model \cite{Wallace}. Here, we take into account the nearest and next-nearest neighbour couplings $\gamma_0\approx 2.7$ eV and $\gamma_n\approx 0.3$ eV \cite{NNN}, respectively, between the carbon atoms on the two triangular sublattices $A$ and $B$ (see Fig. 1a). For pristine graphene, this model yields two bands that touch each other at the same energy in the two inequivalent corners of the hexagonal Brillouin zone (see Fig. 1b), located at $\textbf{K}_{\xi}=(\xi \frac{4\pi}{3a},0)$, often referred to as valleys. Here, $a\approx 2.46\:\mathrm{\AA}$ is the lattice constant of graphene \cite{Lattice_Constant}, and $\xi=\pm 1$ labels the two valleys. For a single electron with momentum $\textbf{p}$ in the vicinity of the valley $\textbf{K}_{\xi}$, an effective low-energy description can be derived,
\begin{align}\label{HamApprox}
&\mathcal{H}_{\textbf{p}}^{{}_{\mathrm{TB},\textbf{K}_{\xi}}}\approx 
\mathcal{H}_{\textbf{p}}^{{}_{0,\textbf{K}_{\xi}}}+
\mathcal{H}_{\textbf{p}}^{{}_{\mathrm{TW},\textbf{K}_{\xi}}}+
\mathcal{H}_{\textbf{p}}^{\mathrm{N}},\\
&\mathcal{H}_{\textbf{p}}^{{}_{0,\textbf{K}_{\xi}}}=
v(\xi  p_x\sigma_x+p_y \sigma_y)+
\mu\sigma_0,\nonumber\\
&\mathcal{H}_{\textbf{p}}^{{}_{\mathrm{TW},\textbf{K}_{\xi}}}=
\frac{v^2}{6\gamma_0}
\left[
\left(p_y^2\!- \!p_x^2\right)\sigma_x\!\!+
2\xi  p_xp_y\sigma_y
\right],\nonumber\\
&\mathcal{H}_{\textbf{p}}^{\mathrm{N}}=-\frac{\gamma_{n}v^2}{\gamma_0^2}
\left(
p_x^2+p_y^2
\right)
\sigma_0.\nonumber
\end{align}
Above, we introduce the Pauli matrices $\sigma_x$, $\sigma_y$ and $\sigma_z$, and the $2\times2$ unit matrix, $\sigma_0$, and write the Hamiltonian in the basis $\left(\phi_{A},\phi_{B}\right)$ of sublattice Bloch states exactly at the centre of the valley. Also, $v=\frac{\sqrt{3}a}{2\hbar}\gamma_0\approx 10^6$ m/s is the Fermi velocity and $\mu$ is the chemical potential.

\begin{figure}\begin{center}
\includegraphics[width=1.0\columnwidth]{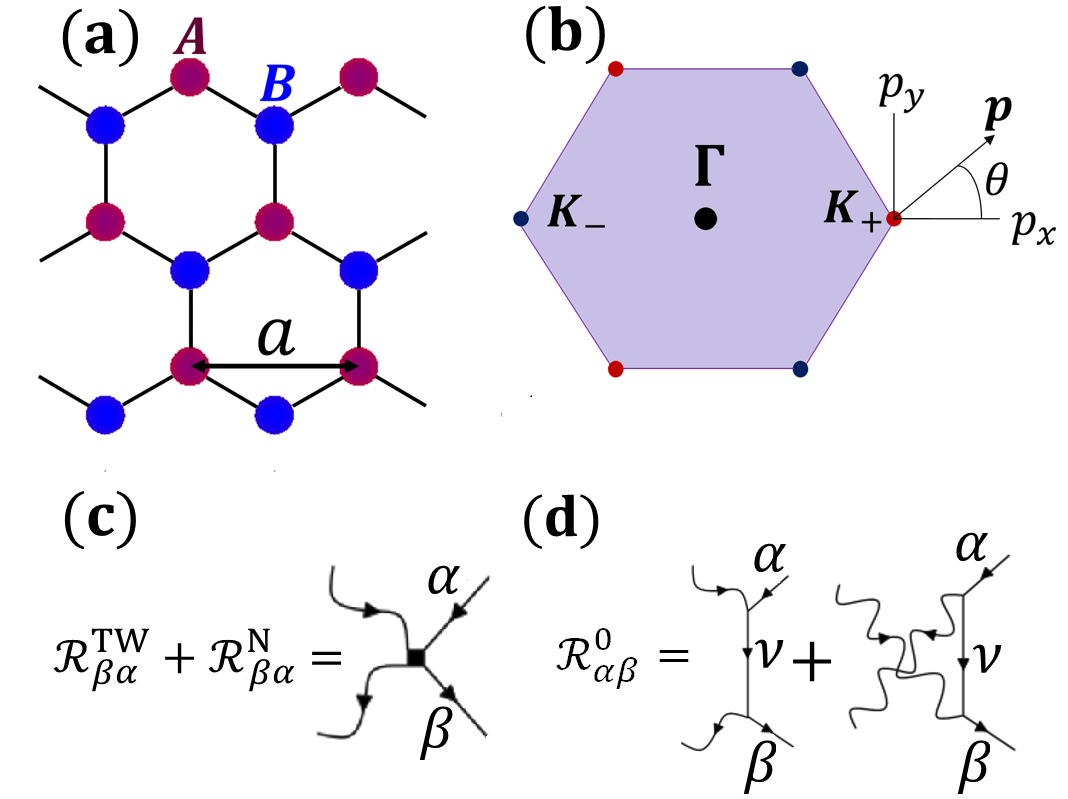}
\caption{  (a) Pictorial representation of the graphene honeycomb lattice with two sublattices, $A$ and $B$. (b) Brillouin zone of monolayer graphene with two inequivalent valleys, denoted by $K_+$ and $K_-$. (c) and (d) Feynman diagrams associated with one-step and two-step processes, respectively. The letters $\alpha$ and $\beta$ denote the band of the electron in the initial and final state, respectively, and $\nu$ corresponds to an intermediate virtual state.}
\end{center}\end{figure}

The term $\mathcal{H}_{\textbf{p}}^{{}_{0,\textbf{K}_{\xi}}}$ in Eq. (\ref{HamApprox}) leads to a linear electronic dispersion $\epsilon_{\alpha,\textbf{p}}=\alpha v p-\mu$, with $\alpha$ denoting the conduction ($\alpha=1$) or valence ($\alpha=-1$) band. Using the eigenstates
\begin{align} 
\hat{\Phi}_{\alpha,\textbf{p}\sigma}^{\textbf{K}_{\xi}}
\left(\textbf{r},t
\right)=
\frac{1}{\sqrt{2}}
\left(
\begin{matrix}
1 \\
\alpha \xi e^{i{\xi} \theta}
\end{matrix}
\right)
e^{i\left(\textbf{p}\textbf{r}-\epsilon_{\alpha,\textbf{p}}t\right)}
\hat{c}_{\alpha, \textbf{p}\sigma }^{\textbf{K}_{\xi}},
\end{align}
and taking into account all electrons, this dominant term in the single-particle Hamiltonian can be written in the diagonal form
\begin{align}\label{HamDiag}
\hat{\mathcal{H}}^0=
\sum_{\xi, \alpha}
\sum_{\textbf{p}, \sigma}
\epsilon_{\alpha,\textbf{p}}
\hat{c}^{\textbf{K}_{\xi}\:\dagger}_{\alpha,\textbf{p}\sigma}
\hat{c}^{\textbf{K}_{\xi}}_{\alpha,\textbf{p}\sigma},
\end{align}
where $\hat{c}_{\alpha, \textbf{p}\sigma}^{\textbf{K}_{\xi}}$ ($\hat{c}_{\alpha, \textbf{p}\sigma}^{\textbf{K}_{\xi}\:\dagger}$) annihilates (creates) an electron in the $\alpha$-band, $\textbf{K}_{\xi}$-valley, with spin $\sigma$ and momentum $\textbf{p}=(p_x,p_y)$ with $\theta=\arctan\left({p_y}/{p_x}\right)$. The other two terms in the right-hand side of  Eq. (\ref{HamApprox}) are small corrections to the Dirac Hamiltonian, $\mathcal{H}_{\textbf{p}}^{{}_{0,\textbf{K}_{\xi}}}$. The trigonal warping, $\mathcal{H}_{\textbf{p}}^{{}_{\mathrm{TW},\textbf{K}_{\xi}}}$, arises from quadratic terms in $\textbf{p}$ generated by the nearest neighbour hopping and lowers the rotational symmetry of the Dirac cones to $C_{3v}$ \cite{Basko2008}. The next-nearest neighbour term, $\mathcal{H}_{\textbf{p}}^{\mathrm{N}}$, introduces asymmetry between the conduction and valence bands \cite{NNN}. 

In order to account for the interaction between photons and electrons in an electronic Raman process, we work with the canonical momentum $\hat{\textbf{P}}={\textbf{p}}-\frac{e}{c}\left(\hat{\textbf{A}}+
\tilde{\hat{\textbf{A}}}\right)$, where $e$ is the electron charge and $\hat{\textbf{A}}$ and $\tilde{\hat{\textbf{A}}}$ are the fields due to the incoming and outgoing light, respectively,
\begin{align}
\hat{\textbf{A}}
\left(\textbf{r},t\right)
&=\frac{\hbar c}{\sqrt{2\Omega}}\left(\textbf{l}e^{i\left(\textbf{q}\cdot\textbf{r}-\Omega t\right)/\hbar}\hat{b}_{\textbf{q},q_z,\textbf{l}}+h.c.\right),\nonumber
\\
\tilde{\hat{\textbf{A}}}
\left(\textbf{r},t\right)
&=\frac{\hbar c}{\sqrt{2\tilde{\Omega}}}\left(\tilde{\textbf{l}}e^{i\left(
\tilde{\textbf{q}}\cdot\textbf{r}-\tilde{\Omega} t\right)/\hbar}\hat{b}_{\tilde{\textbf{q}},\tilde{q}_z,\tilde{\textbf{l}}}+h.c.\right).\nonumber
\end{align}
Here, the operator $\hat{b}_{\textbf{q},q_z,\textbf{l}}$ ($\hat{b}_{\tilde{\textbf{q}},\tilde{q}_z,\tilde{\textbf{l}}}^{\dagger}$) annihilates (creates) a photon with in-plane momentum ${\textbf{q}}$ ($\tilde{\textbf{q}}$), out-of-plane momentum ${q}_z$ ($\tilde{q}_z$), polarization $\textbf{l}=(l_x,l_y)$ [$\tilde{\textbf{l}}=({\tilde{l}}_x,{\tilde{l}}_y)$] and energy $\Omega$ ($\tilde{\Omega}$) of the incident (scattered) photon. Keeping contributions up to the second order in the vector potential yields the interaction part
\begin{align}\label{LightInt}
{\mathcal{H}}^{\mathrm{light}}\left(\textbf{r},t\right)=
&-\frac{e}{c}\frac{\partial {\mathcal{H}}^{\mathrm{TB}}}{\partial \textbf{p}}
\left(\hat{\textbf{A}}\left(\textbf{r},t\right)+
\tilde{\hat{\textbf{A}}}\left(\textbf{r},t\right)\right)\\
&+\frac{e^2}{2c^2}  \sum_{n,m}
\frac{\partial^2 {\mathcal{H}}^{{\mathrm{TB}}}}{\partial p_n \partial p_m}
\hat{{A}}_n\left(\textbf{r},t\right)
\tilde{\hat{{A}}}_m\left(\textbf{r},t\right).\nonumber
\end{align}
The two terms in the expression above correspond to two different contributions to the scattering amplitude of the electronic Raman process.  The second term gives rise to a one-step process, the usual contact interaction \cite{Light_Scattering,Wolff}, illustrated by the Feynman diagram in Fig. 1c, in which an incoming photon scatters inelastically on an electron passing to it instantly energy equal to the Raman shift $\omega=\Omega-\tilde{\Omega}$. Because of the second derivative in momentum, only the trigonal warping and electron-hole asymmetry terms generate one-step processes. For a Raman process that scatters an electron from an initial state $\hat{\Phi}_{\alpha,\textbf{p}\sigma}^{\textbf{K}_{\xi}}$ into a final state $\hat{\Phi}_{\beta,\textbf{p}'\sigma}^{\textbf{K}_{\xi}}$, the corresponding scattering amplitudes, $\hat{\mathcal{R}}^{\mathrm{TW}}_{\beta\alpha}$ and $\hat{\mathcal{R}}^{\mathrm{N}}_{\beta\alpha}$, are obtained by taking into account scattering at any time $t_1\in(-T,T)$, where $2T$ is the exposure time,
\begin{widetext}
\begin{align}\label{RamanContact}
\hat{\mathcal{R}}^{TW}_{\beta\alpha}\!=\!
&\frac{e^2\hbar^2v^2}{6\gamma_02\sqrt{\Omega\tilde{\Omega}}}
\int_{-T}^{T}\!
\frac{\mathrm{d}t_1 }{i\hbar} 
\hat{\Phi}_{\beta, \textbf{p}'\sigma }^{\textbf{K}_{\xi} \: \dagger}\!
\left(\textbf{r},t_1\right)
e^{-\frac{i}{\hbar}\left(
\tilde{\textbf{q}}\textbf{r}-{\tilde{\Omega}t_1}
\right)}
\hat{b}^{\dagger}_{\tilde{\textbf{q}},\tilde{q}_z,\tilde{\textbf{l}}}\left[
\!\left(
{l}_y\tilde{l}_y^*\!-\!l_x\tilde{l}_x^*
\right)\!
\sigma_x\!\!+\!\xi\!
\left(
l_x\tilde{l}_y^*\!+\!{l}_x\tilde{l}_y^*
\right)\!
\sigma_y
\right]\!\hat{b}^{}_{{\textbf{q}},q_z,{\textbf{l}}}
 e^{\frac{i}{\hbar}\left(
{\textbf{q}}\textbf{r}-\Omega t_1
\right)}
\hat{\Phi}_{\alpha,\textbf{p}\sigma}^{\textbf{K}_{\xi}}
\left(
\textbf{r},t_1
\right),\\
\hat{\mathcal{R}}^{\mathrm{N}}_{\beta\alpha}=
&\frac{-\gamma_ne^2\hbar^2v^2}{\gamma_0^22\sqrt{\Omega\tilde{\Omega}}}
\int_{-T}^{T} \!
\frac{\mathrm{d}t_1 }{i\hbar}
\hat{\Phi}_{\beta,\textbf{p}'\sigma }
^{\textbf{K}_{\xi} \: \dagger}\!
\left(\textbf{r},t_1\right)
e^{-\frac{i}{\hbar}\left(
\tilde{\textbf{q}}\textbf{r}\!-\!{\tilde{\Omega}t_1}
\right)}
\hat{b}^{\dagger}_{\tilde{\textbf{q}},\tilde{q}_z,\tilde{\textbf{l}}}
\left[\left(
l_x\tilde{l}_x^*\!+\!l_x\tilde{l}_y^*
\right)\sigma_0\right]
\hat{b}^{}_{{\textbf{q}},q_z,{\textbf{l}}}e^{\frac{i}{\hbar}\left(
{\textbf{q}}\textbf{r}-\Omega t_1
\right)}
\hat{\Phi}_{\alpha,\textbf{p}\sigma}^{\textbf{K}_{\xi}}
\left(\textbf{r},t_1\right).\nonumber
\end{align}
\end{widetext}

In turn, the two-step processes generated by the first term in ${\mathcal{H}}_{\textbf{p}}^{\mathrm{light}}$, involve an intermediate virtual electronic state $\hat{\Phi}_{\nu,\textbf{p}''\sigma}^{\textbf{K}_{\xi}}$ in the band $\nu$, created at $t_1$ and annihilated at $t_2$. Because this term contains the first derivative with respect to momentum, the Dirac term $\hat{\mathcal{H}}^{0,K_{\xi}}_{\textbf{p}}$ in Eq. (\ref{HamApprox}), is involved. The scattering amplitude for a two-step process between bands $\alpha$ and $\beta$ is:
\begin{widetext}
\begin{align}\label{RamanLinTime}
\hat{\mathcal{R}}^0_{\beta\alpha}=
-&\frac{e^2\hbar^2v^2}{2\sqrt{\Omega\tilde{\Omega}}}\!
\int_{-T}^{T}
\int_{-T}^{t_2}\!
\frac{\mathrm{d}t_2 \mathrm{d}t_1}{\hbar^2}
\hat{\Phi}_{\beta,\textbf{p}'\sigma }
^{\textbf{K}_{\xi} \: \dagger}\left(
\textbf{r},t_2
\right)e^{-\frac{i}{\hbar}(\tilde{\textbf{q}}\textbf{r}-\tilde{\Omega} t_2)}\hat{b}^{\dagger}_{{\tilde{\textbf{q}}},\tilde{q}_z,\tilde{\textbf{l}}}
\left(
\xi\tilde{l}_x^*\sigma_x+\tilde{l}_y^*\sigma_y
\right)\\
&\times\sum_{\nu}
\hat{\Phi}_{\nu, \textbf{p}''\sigma }^{\textbf{K}_{\xi}}
\left(\textbf{r},t_2\right)
\hat{\Phi}_{\nu,\textbf{p}''\sigma }^{\textbf{K}_{\xi} \: \dagger}
\left(\textbf{r},t_1\right)\nonumber
\left(\xi{l}_x\sigma_x+{l}_y\sigma_y\right)
\hat{b}^{}_{\textbf{q},q_z,{\textbf{l}}}\nonumber
e^{\frac{i}{\hbar}(\textbf{q}\textbf{r}-\Omega t_1)}
\hat{\Phi}_{\alpha,\textbf{p}\sigma}^{\textbf{K}_{\xi}}
\left(
\textbf{r},t_1
\right)
\nonumber\\
-&\frac{e^2\hbar^2v^2}{2\sqrt{\Omega\tilde{\Omega}}}\!
\int_{-T}^{T}
\int_{-T}^{t_2}
\frac{\mathrm{d}t_2\mathrm{d}t_1 }{\hbar^2}
\hat{\Phi}_{\beta,  \textbf{p}'\sigma}^{\textbf{K}_{\xi} \: \dagger}
\left(
\textbf{r},t_2
\right)
e^{\frac{i}{\hbar}({\textbf{q}}\textbf{r}-{\Omega} t_2)}\nonumber
\left(\xi{l}_x\sigma_x+{l}_y\sigma_y\right)
\hat{b}_{\textbf{q},q_z,{\textbf{l}}}\nonumber\\
&\times\sum_{\nu}
\hat{\Phi}_{\nu, \textbf{p}''\sigma }^{\textbf{K}_{\xi}}
\left(\textbf{r},t_2\right)
\hat{\Phi}_{\nu,\textbf{p}''\sigma }^{\textbf{K}_{\xi} \: \dagger}
\left(\textbf{r},t_1\right)\nonumber
\hat{b}^{\dagger}_{{\tilde{\textbf{q}}},\tilde{q}_z,\tilde{\textbf{l}}}
\left(\xi\tilde{l}_x^*\sigma_x+\tilde{l}_y^*\sigma_y\right)
\nonumber
e^{-\frac{i}{\hbar}(\tilde{\textbf{q}}\textbf{r}-\tilde{\Omega} t_1)}
\hat{\Phi}_{\alpha, \textbf{p}\sigma}^{\textbf{K}_{\xi}}
\left(\textbf{r},t_1\right),\nonumber
\end{align}
\end{widetext}
where the term in the first (second) two lines, corresponds to processes in which the photon is absorbed (emitted) at a time $t_1$ and emitted (absorbed) at a time $t_2$, and is expressed by the first (second) diagram in Fig. 1d. In Eq. (\ref{RamanLinTime}), integration over $t_1$ in both terms generates the factors $(\pm\bar{\Omega}-\epsilon_{\nu})^{-1}$, where
$\bar{\Omega}=(\Omega+\tilde{\Omega})/{2}$. Assuming that the energy of electronic Raman excitations, $\omega$, is significantly smaller than the energy of the incoming and outgoing photons, we can expand these factors in powers of $\epsilon_{\nu}/\bar{\Omega}$. In these expansions, we only retain the leading 0-th order term which, in contrast to many materials \cite{Wolff}, does not vanish for graphene \cite{Kashuba_Falko1, Theory_ERS_BLG}.

Both one- and two-step processes result in the creation of an electron-hole pair in the electronic bands of monolayer graphene, an excited state, $\Ket{\mathrm{exc}}$, over the ground state of the system, $\Ket{\mathrm{GS}}$. In the absence of superconductivity, such ground state is the Fermi sea of electronic states filled up to the chemical potential, $\Ket{\mathrm{GS}}=\prod_{\epsilon_{\alpha,\textbf{p}}<\mu}
\hat{c}^{\textbf{K}_{\xi}\:\dagger}_{\beta,\textbf{p}\sigma}\Ket{\mathrm{vac}}
$, where $\Ket{\mathrm{vac}}$ represents the electronic vacuum. Then, the excited state takes the form $\Ket{\mathrm{exc}}=
\hat{c}^{\textbf{K}_{\xi}\:\dagger}_{\beta,\textbf{p}'\sigma}
\hat{c}^{\textbf{K}_{\xi}}_{\alpha,\textbf{p}\sigma}\Ket{\mathrm{GS}}$. 

For pristine graphene, the chemical potential lies at the Dirac points, and the initial and final states must belong to the valence ($\alpha=-1$) and conduction ($\beta=+1$) band, respectively. More generally, for doped graphene, the initial and final state could also belong to the same band. Here, we consider strongly $n$-doped graphene and use Fermi's golden rule to relate the total scattering amplitude,
\begin{equation}
\hat{\mathcal{R}}_{+\alpha}=\hat{\mathcal{R}}^0_{+\alpha}
\!+\!\hat{\mathcal{R}}^{\mathrm{TW}}_{+\alpha}
\!+\!\hat{\mathcal{R}}^{\mathrm{N}}_{+\alpha},\nonumber
\end{equation}
to the angle-resolved probability ${w(\omega,\tilde{\textbf{q}}})$ of an ERS process,
\begin{align}\label{FGR}
w(\omega,\tilde{\textbf{q}})
=\!\!
\lim_{T\rightarrow\infty}\!\sum_{\alpha,\Ket{\mathrm{exc}}}\int
\frac{\mathrm{d}\textbf{p}}{2\pi\hbar^3}
|\!
\Bra{\mathrm{exc}}
\sum_{\xi,\sigma}\!
\hat{\mathcal{R}}_{+\alpha}
\Ket{\mathrm{GS}}\! |^2.
\end{align}
In the single-particle picture described by Eq. (1), spins and valleys are uncoupled and remain good quantum numbers that uniquely identify the excited state. Therefore, the sum over these quantum numbers in the equation above results simply in a factor of four in the overall probability. The integration over time in Eqs. (\ref{RamanContact}) and (\ref{RamanLinTime}), together with the limit $T\rightarrow\infty$ in Eq. (\ref{FGR}), leads in the usual way \cite{QuantumMechanics} to the conservation of energy, $\omega=\epsilon_{+,\textbf{p}'}-\epsilon_{\alpha,\textbf{p}}$. In turn, evaluation of the scattering matrix element imposes conservation of momentum $\textbf{p}'=\textbf{p}+\tilde{\textbf{q}}-\textbf{q}$, which forces momentum of the final electronic state $\textbf{p}'$ to be on a circle of radius $|\tilde{\textbf{q}}-\textbf{q}|$ in the momentum space around $\textbf{p}$. However, for Fermi level $\mu\sim0.1$ eV, we can neglect the momentum transfer from the photon to the electron and consider all electron excitations to be vertical, $\textbf{p}'=\textbf{p}$. As a result, the excited state $\Ket{\mathrm{exc}}=\hat{c}^{\textbf{K}_{\xi}\:\dagger}_{\alpha,\textbf{p}\sigma}
\hat{c}^{\textbf{K}_{\xi}}_{\alpha,\textbf{p}\sigma}\Ket{\mathrm{GS}}=\Theta(\mu-\epsilon_{\alpha,\textbf{p}})\Ket{\mathrm{GS}}$, where $\Theta(x)$ is the Heaviside function, cannot give rise to a Raman shift, and only the inter-band excitations, here $\Ket{\mathrm{exc}}=\hat{c}^{\textbf{K}_{\xi}\:\dagger}_{+,\textbf{p}\sigma}
\hat{c}^{\textbf{K}_{\xi}}_{-,\textbf{p}\sigma}\Ket{\mathrm{GS}}$, contribute to the electronic Raman spectrum \cite{Theory_ERS_MLG},
\begin{align}\label{Winter}
&{w}(\omega)=\frac{\hbar e^4v^2}{\Omega^2}
\left[
\frac{\Xi_{\mathrm{s}}}{\Omega^2}
+
\frac{\Xi_{\mathrm{o}}}{2\left(6\gamma_0\right)^2}
\right]
\omega\:\Theta(\omega-2\mu),\\
&\Xi_{s}=|
\textbf{l}\times\tilde{\textbf{l}}^*
|^2, \qquad
\Xi_{\mathrm{o}}=1+({\textbf{l}}\times{\textbf{l}}^*)(\tilde{\textbf{l}}\times\tilde{\textbf{l}}^*).\nonumber
\end{align}
Above, the first term with polarization factor $\Xi_{\mathrm{s}}$ describes the contribution from in/out photons that carry the same circular polarization. The second term, accompanied by the polarization factor $\Xi_{\mathrm{o}}$, represents processes with the opposite circular polarization of the incoming and scattered light. In terms of linear polarization, $\Xi_{\mathrm{s}}$ corresponds to crossed polarization of in/out photons, while $\Xi_{\mathrm{o}}$ is 1 independently of the linear polarization of the incident and scattered light. The values of the polarization factors, depending on the relative polarization of the in/out photons, are listed in Table \ref{PolarizationTable}. The Heaviside function captures the Pauli blocking of excitations with Raman shift $\omega<2\mu$ in graphene doped to Fermi level $\mu\neq 0$.

For inter-band transitions, the dominant Raman signal is due to excitations that conserve angular momentum. Also, notice that the electron-hole asymmetry scattering amplitude, $\hat{\mathcal{R}}_{+-}^{\mathrm{N}}$, does not contribute to the Raman response because its corresponding electronic Hamiltonian, $\mathcal{H}^{\mathrm{N}}_{\textbf{p}}$, is proportional to the unit matrix, and hence only couples states in the same band. 

After integrating over all directions of propagation of the scattered photons, we obtain the spectral density of the angle-integrated Raman signal \cite{Theory_ERS_MLG},
\begin{align}
g(\omega)=&\int\int
\frac{\mathrm{d}\textbf{q}\mathrm{d}q_z}{\left(2\pi\hbar\right)^3}
\mathrm{w}(\omega)\,
\delta\!
\left(
\tilde{\Omega}-c\sqrt{\tilde{\textbf{q}}-\tilde{q}_z}
\right)\\
=&\frac{e^4v^2}{4\hbar^2\pi^2c^4}
\left[
\frac{\Xi_{\mathrm{s}}}{\Omega^2}
+
\frac{\Xi_{\mathrm{o}}}{2\left(6\gamma_0\right)^2}
\right]
\omega\:\Theta(\omega-2\mu).\nonumber
\end{align}
In turn, the spectral density allows us to estimate the quantum efficiency, defined as the flux of outgoing photons to the incoming photons, by integrating over all possible energy shifts, $I=\int d\omega g(\omega)$. 

The linear dependence of $g(\omega)$ on $\omega$, a consequence of the linear electronic density of states in monolayer graphene, was confirmed experimentally together with the change in intensity depending on crossed or parallel linear polarization of incident/detected photons \cite{Exp_ERS_graphene_1}.

\section{\label{sec:level3} ERS in superconducting graphene}
We introduce superconductivity in graphene by considering a pairing interaction between two electrons with opposite spins and momenta that gives rise to an isotropic gap in the dispersion relation, $
\hat{\mathcal{H}}^{\mathrm{pair}}=-g\hat{\Psi}^{\dagger}\hat{\Psi}
$, where $\hat{\Psi}=\sum_{\textbf{k}}\hat{c}_{-\textbf{k}\downarrow}
\hat{c}_{\textbf{k}\uparrow}$ is the field generated by the coupled electrons and $g$ is the strength of the coupling \cite{BCS}. However, in our low energy description, an electronic state with momentum $\textbf{k}$, measured from the centre of the Brillouin zone, is mapped on a state in the valley $\textbf{K}_{\xi}$ and momentum $\textbf{p}$ measured from the centre of that valley. The presence of one more good quantum number, the valley-index $\xi$, allows two possible configurations for a BCS-like field in graphene, so that we substitute for $\hat{\Psi}$ a new field, $\hat{\Psi}_s$, such that
\begin{align}\label{PossibleStates}
&\hat{\mathcal{H}}^{\mathrm{pair}}=
-g\hat{\Psi}^{\dagger}_s\hat{\Psi}_s,\\
&\hat{\Psi}_s=
\sum_{\alpha,\textbf{p}}\frac{1}{\sqrt{2}}
\left(
\hat{c}_{\alpha, -\textbf{p}\downarrow}^{\textbf{K}_- }
\hat{c}_{\alpha, \textbf{p}\uparrow}^{\textbf{K}_+ }+s
\hat{c}_{\alpha, -\textbf{p}\downarrow}^{\textbf{K}_+ }
\hat{c}_{\alpha, \textbf{p}\uparrow}^{\textbf{K}_- }
\right).\nonumber
\end{align}
In the equation above, the index $s=\pm1$ distinguishes between two possible wave functions: the $\hat{\Psi}_{+}$ singlet state is antisymmetric and the $\hat{\Psi}_{-}$ triplet state symmetric under the inversion of all the spins. The spatial parts of the singlet and triplet wave functions belong to the $A_1$ and $B_1$ irreducible representations of the symmetry group $C_{6v}''$, consisting of the point group of the graphene crystal $C_{6v}$ and primitive translations \cite{Basko2008}. 

We employ the mean-field approximation \cite{FetterWallecka}, 
\begin{equation}
\hat{\Psi}^{\dagger}_s\hat{\Psi}_s\approx \Braket{\hat{\Psi}_s}^{\dagger}\hat{\Psi}_s+\hat{\Psi}_s^{\dagger}\Braket{\hat{\Psi}_s}-\Braket{\hat{\Psi}_s^{\dagger}}\Braket{\hat{\Psi}_s},\nonumber
\end{equation}
and introduce the order parameter $\Delta=g\Braket{\hat{\Psi}_s}$. In order to determine the quasiparticle dispersion, we add the pairing interaction to the dominant linear term in the single-particle Hamiltonian and neglect all other contributions.  
\begin{table}[b!]
\centering
\caption{Values of the polarization factors used in the text depending on the relative polarization of the incident and scattered light.}\label{PolarizationTable}
\begin{tabular}{| c | c c | c c |} 
 \hline 
  $\quad\,$ & $\textbf{Circularly}$& $\textbf{polarised}$ &
  $\textbf{Linearly}$ & $\textbf{polarised}$\\
 \hline
   &Same &  Opposite & Parallel &  Crossed \\ [0.2ex] 
 \hline
   $\Xi_o$&0 & 2& 1 & 1 \\ [0.1ex]
   \hline
 $\Xi_s$&  1 & 0& 0 & 1 \\ [0.1ex]
 \hline
  $\Xi_s'$&  1 & 0 & 1 & 0 \\ [0.1ex]
 \hline
 \end{tabular}
\end{table}Making use of the diagonal form of the linear term, Eq. (\ref{HamDiag}), we write the Hamiltonian of the electrons in superconducting graphene, in the basis $\hat{A}_{\alpha,\textbf{p}}^{\dagger}=\left(
\hat{c}_{\alpha,\textbf{p},\uparrow}^{\textbf{K}_+\:\dagger}\:
{\hat{c}_{\alpha,-\textbf{p},\downarrow}^{\textbf{K}_-}}\:
\hat{c}_{\alpha,\textbf{p},\uparrow}^{\textbf{K}_-\: \dagger}\:
\hat{c}_{\alpha,-\textbf{p},\downarrow}^{\textbf{K}_+}\right)$,
\begin{align}\label{A1B1Ham}
&{\hat{\mathcal{H}}}^{\mathrm{0}}+
{\hat{\mathcal{H}}}^{\mathrm{pair}}=
\sum_{\alpha,\textbf{p}}\hat{A}_{\alpha,\textbf{p}}^{\dagger}
\mathcal{H}^{s}_{\textbf{p}}
\hat{A}_{\alpha,\textbf{p}},\\
&{{\mathcal{H}}}^{s}_{\textbf{p}}=
\left(
\begin{matrix}
\epsilon_{{}_{\alpha,\textbf{p}}}\sigma_z-\Delta\sigma_x&0\\
0&\epsilon_{{}_{\alpha,\textbf{p}}}\sigma_z-s\Delta\sigma_x
\end{matrix}
\right).\nonumber
\end{align}
This Hamiltonian is diagonalised by a Bogoliubov transformation $\mathcal{U}_{\textbf{p}}^{s}\hat{\mathcal{H}}_{\textbf{p}}^s\mathcal{U}_{\textbf{p}}^{s\:\dagger}$, where
\begin{align}\label{transformation}
\mathcal{U}_{\textbf{p}}^{s}=\left(
\begin{matrix}
u_{{}_{\alpha,\textbf{p}}}\sigma_0+v_{{}_{\alpha,\textbf{p}}}i \sigma_y & 
0\\
0 &
u_{{}_{\alpha,\textbf{p}}}\sigma_0+sv_{{}_{\alpha,\textbf{p}}}i \sigma_y 
\end{matrix}
\right),\\
u_{\alpha,\textbf{p}}=
\sqrt{\frac{\varepsilon_{\alpha,\textbf{p}}+
\epsilon_{\alpha,\textbf{p}}}{2\varepsilon_{\alpha,\textbf{p}}}},\quad
v_{\alpha,\textbf{p}}=\sqrt{\frac{\varepsilon_{\alpha,\textbf{p}}-
\epsilon_{\alpha,\textbf{p}}}{2\varepsilon_{\alpha,\textbf{p}}}}.\nonumber
\end{align}
The eigenvectors, in turn, allow us to define the Bogoliubov quasiparticles
\begin{equation}\label{quasiparticles}
\mathcal{U}_{\textbf{p}}^{s\:\dagger}
\hat{A}_{\alpha,\textbf{p}}\!=\!
\left(\begin{matrix}
\hat{\gamma}_{\alpha \textbf{p}\uparrow}^{K_{+}}\vspace{0.8mm}\\
{\hat{\gamma}_{\alpha \textbf{p}\downarrow}^{K_{-}\:\dagger}}\vspace{0.8mm}\\
\hat{\gamma}_{\alpha \textbf{p}\uparrow}^{K_-}\vspace{0.8mm}\\
{\hat{\gamma}_{\alpha \textbf{p}\downarrow}^{K_+\:\dagger}}\vspace{0.8mm}\\
\end{matrix}\right)\!=\!
\left(\begin{matrix}
u_{\alpha,\textbf{p}}
\hat{c}_{\alpha \textbf{p}\uparrow}^{K_{+}}-
v_{\alpha,\textbf{p}}
{\hat{c}_{\alpha \textbf{p}\downarrow}^{K_{-}\:\dagger}}\vspace{0.8mm}\\
u_{\alpha,\textbf{p}}
{\hat{c}_{\alpha \textbf{p}\downarrow}^{K_{-}\:\dagger}}+
v_{\alpha,\textbf{p}}
\hat{c}_{\alpha \textbf{p}\uparrow}^{K_{+}}\vspace{0.8mm}\\
u_{\alpha,\textbf{p}}
\hat{c}_{\alpha \textbf{p}\uparrow}^{K_{-}}-
sv_{\alpha,\textbf{p}}
{\hat{c}_{\alpha \textbf{p}\downarrow}^{K_{+}\:\dagger}}\vspace{0.8mm}\\
u_{\alpha,\textbf{p}}
{\hat{c}_{\alpha \textbf{p}\downarrow}^{K_{+}\:\dagger}}+
sv_{\alpha,\textbf{p}}
\hat{c}_{\alpha \textbf{p}\uparrow}^{K_{-}}\vspace{0.8mm}\\
\end{matrix}\right).
\end{equation}
Their dispersion, $\varepsilon_{\alpha,\textbf{p}}=\sqrt{\epsilon_{\alpha,\textbf{p}}^2+\Delta^2}$, is four-fold degenerate and features a superconducting gap at the position of the chemical potential. The ground state of the system consists of a quasiparticle vacuum which satisfies by definition, $\hat{\gamma}_{\alpha,\textbf{p}\sigma}^{K_{\xi}}\Ket{\mathrm{GS}}=0$. Notice that, for $\Delta\rightarrow0$, the Bogoliubov coefficient $v_{\alpha,\textbf{p}}$ ($u_{\alpha,\textbf{p}}$) takes the value one (zero), below the Fermi surface, and zero (one) above, turning the superconducting ground state into the Fermi sea of non-interacting electrons.

\begin{figure}
\begin{center}
\includegraphics[width=1.0\columnwidth]{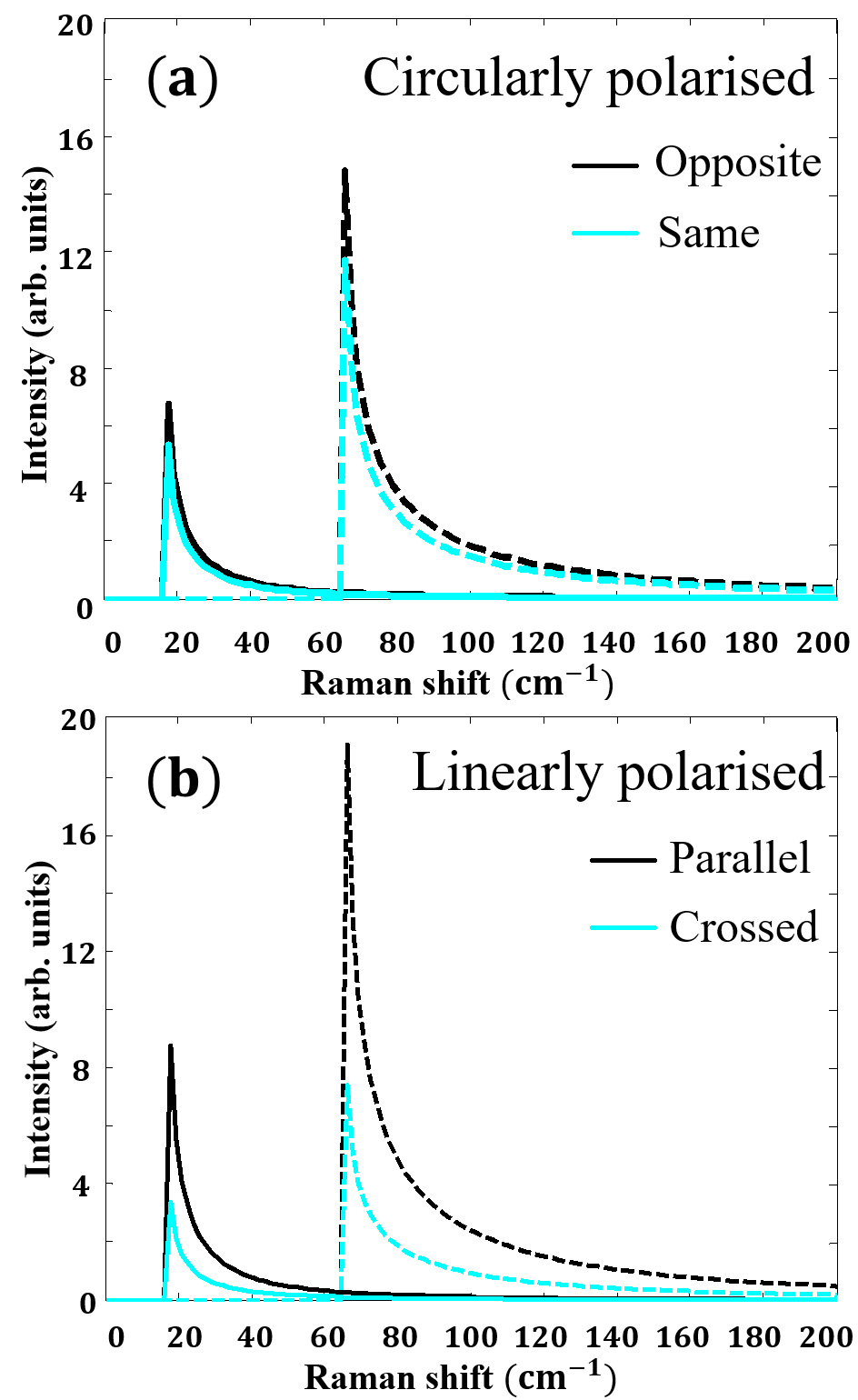}
\caption{ The low-energy electronic contribution to the Raman spectrum of superconducting graphene with chemical potential $\mu=150$ meV, for incoming photon energy $\Omega=1$ eV and (a) circular and (b) linear polarization of the incoming/scattered light. The solid and dashed lines correspond to the order parameter $\Delta=1$ meV and $\Delta=4$ meV, respectively.}
\label{RESULTS}\end{center}\end{figure}

In order to use Eq. (\ref{FGR}) for the calculation of the angle-resolved scattering probability $w(\omega)$ for superconducting monolayer graphene, we need to express the pairs of electron operators $\hat{c}^{\textbf{K}_{\xi}\:\dagger}_{\beta,\textbf{p}\sigma}\hat{c}^{\textbf{K}_{\xi}}_{\alpha,\textbf{p}\sigma}$, and hence the scattering amplitudes in Eqs.(\ref{RamanContact}) and (\ref{RamanLinTime}), in terms of pairs of operators involving $\hat{\gamma}^{K_{\xi}}_{\alpha,\textbf{p}\sigma}$ and $\hat{\gamma}^{K_{\xi}\:\dagger}_{\alpha,\textbf{p}\sigma}$. Because the ground state is a quasiparticle vacuum, terms that first annihilate a quasiparticle give zero when acting on a ground state, and the remaining terms yield
\begin{align}\label{ExcitedStates}
\hat{c}^{\textbf{K}_{\xi}\:\dagger}_{+,\textbf{p}\sigma}
\hat{c}^{\textbf{K}_{\xi}}_{\alpha,\textbf{p}\sigma}
\Ket{\mathrm{GS}}
&=v_{{}_{+,\textbf{p}}}v_{{}_{\alpha,\textbf{p}}}
\hat{\gamma}_{+,\textbf{p}-\sigma}^{K_{-\xi}}
\hat{\gamma}_{\alpha,\textbf{p}-\sigma}^{K_{-\xi}\:\dagger}
\Ket{\mathrm{GS}}\\
&+u_{{}_{+,\textbf{p}}}
v_{{}_{\alpha,\textbf{p}}}
\hat{\gamma}_{+,\textbf{p}-\sigma}^{K_{\xi}\:\dagger}
\hat{\gamma}_{\alpha,-\textbf{p}-\sigma}^{K_{-\xi}\:\dagger}
\Ket{\mathrm{GS}}.\nonumber
\end{align}
For $\alpha=-1$, the first term in the right-hand side of the equation above vanishes and the second one generates an inter-band feature similar to that described in the previous section. In turn, for $\alpha=+1$, the first term creates and annihilates the same quasiparticle, leading to no Raman shift. The second term, which appears purely because of the superconducting pairing, generates two quasiparticles yielding one of two possible excited states, $\Ket{\mathrm{exc}}=
\hat{\gamma}_{+,\textbf{p}\uparrow}^{K_{\xi}\:\dagger}
\hat{\gamma}_{+,-\textbf{p}\downarrow}^{K_{-\xi}\:\dagger}
\Ket{\mathrm{GS}}$, with excitation energy $2\varepsilon_{+,\textbf{p}}$, thus leading to an intra-band ERS feature. Using Eq. (\ref{FGR}), we obtain the angle-resolved and angle-integrated Raman probabilities,
\begin{align}\label{FGR2}
&{w}(\omega)\!= \!\frac{8\hbar e^4 v^2 \mu}{\Omega^2}\!
\left[
\left(\frac{\mu}{\Omega^2}\!-\!\frac{\gamma_{\mathrm{n}}}{2\gamma_0^2}
\right)^2\!\!\Xi'_{\mathrm{s}}
\!+\!
\frac{\mu^2}{2\Omega^4}\Xi_{\mathrm{o}}
\right]\!f(\omega),
\\
&g(\omega)\!=\!\frac{2e^4 v^2 \mu}{\pi^2\hbar^2c^4}\!
\left[
\left(\frac{\mu}{\Omega^2}\!-\!\frac{\gamma_{\mathrm{n}}}{2\gamma_0^2}
\right)^2\!\!\Xi'_{\mathrm{s}}
\!+\!
\frac{\mu^2}{2\Omega^4}\Xi_{\mathrm{o}}
\right]\!f(\omega),\nonumber
\\
&f(\omega)=
\frac{4\Delta^2}{\omega\sqrt{\omega^2-4\Delta^2}}, \qquad
\Xi'_{\mathrm{s}}=|\textbf{l}\cdot\tilde{\textbf{l}}^*|^2.
\nonumber
\end{align}
In equations above, the new polarization factor $\Xi'_{\mathrm{s}}$ describes a contribution of processes with the same circular polarization or parallel linear polarization of the in/out photons, in contrast to the factor $\Xi_{\mathrm{s}}$ in Eq. (\ref{FGR}) which selects crossed in/out linear polarization (see Table \ref{PolarizationTable}). Notice that the next-nearest neighbour hopping contributes to the new Raman feature. The trigonal warping, however, does not because the scattering amplitudes for two processes with reversed valley indices are opposite and cancel out.

We show the superconductivity-induced intra-band feature in Fig. \ref{RESULTS}, for a Fermi level shifted $\mu=150$ meV above the Dirac point, energy of the incoming photons $\Omega=1$ eV and values of the order parameter $\Delta=1$ meV  and $\Delta=4$ meV (solid and dashed lines, respectively). In the panel (a), we show the comparison between detecting scattered photons with circular polarization the same or oppossite to the incoming photons. In the panel (b), we present a similar comparison for the case of linear polarization. In contrast to the inter-band ERS feature, for the superconductivity-induced signal the only difference between processes that conserve or change angular momentum of the quasiparticle sea is due to the next-nearest neighbour coupling. For strongly doped graphene, $\frac{\gamma_{\mathrm{n}}}{\gamma_0^2}\ll\frac{\mu}{\Omega^2}$, we expect the intra-band Raman response to be insensitive to the circular polarization of in/out photons. In comparison, for the case of linear polarization, the signal due to scattered light with polarization parallel to that of the incident light is twice as strong as for crossed polarization of in/out photons.

In general, the asymmetric shape of the new peak is governed by the density of states close to the gap. Hence, the intensity of the peak formally diverges at the Raman shift $\omega=2\Delta$. However, we discuss its height by looking at the intensity at the Raman shift $\omega=2\Delta+\delta$, where $\delta$ determines the distance from the divergence. The functional dependence
\begin{equation}
f(2\Delta+\delta)=\frac{4\Delta^2}{(2\Delta+\delta)\sqrt{(2\Delta+\delta)^2-4\Delta^2}}\approx \sqrt{\frac{\Delta}{\delta}}\nonumber
\end{equation}
shows that the intensity of the superconductivity-induced peak scales with the square root of $\Delta$. 

For $\frac{\gamma_{\mathrm{n}}}{\gamma_0^2}\ll\frac{\mu}{\Omega^2}$, the overall signal is proportional to the third power of the chemical potential $\mu$ and inversely proportional to the sixth power of the incoming photon energy $\Omega$. We estimate the quantum efficiency of the ERS peak $I=\int_{2\Delta}^{\infty}\mathrm{d}\omega g(\omega)\sim\left(\frac{e^2}{\hbar c}\frac{v}{c}\right)^2\!\frac{\Delta \mu^3}{\Omega^4}\sim 10^{-14}$.

\section{Summary}
We have investigated the electronic Raman spectrum of highly doped and superconducting graphene with an isotropic superconducting order parameter within a single valley. We predict a Raman peak at a shift of twice the superconducting order parameter, $\omega=2\Delta$. Modern Raman spectroscopy measures shifts $\omega\sim1$ meV with resolution $\sim0.1$ meV \cite{RamanResolution}, suggesting that features presented in this paper, corresponding to experimentally observed gaps \cite{Li-Decorated}, might be observable. The quantum efficiency of the predicted peak, $I\sim10^{-14}$, is three orders of magnitude smaller than the well-known graphene phonon-induced G-peak \cite{Gpeak}, and two orders of magnitude smaller than the experimentally observed electronic Raman features due to inter-Landau-level transitions \cite{Theory_ERS_MLG,ExpConf, ExpConf2,ExpConf3}. However, because of the Pauli blocking of excitations for $\omega<2\mu$, the superconductivity-induced peak is the only low-energy electronic feature in the Raman spectrum.

\section{\label{sec:level5}Acknowledgments}
This work has been supported by the UK Engineering and Physical Sciences Research Council (EPSRC) through the Centre for Doctoral Training in Condensed Matter Physics (CDT-CMP), Grant No. EP/L015544/1, as well as EPSRC Grant EP/N010345/1, the European Graphene Flagship project and Lloyd's Register Foundation Nanotechnology Programme.

\end{document}